\documentclass[a4paper]{ESASPCS13Style_astroph} 
\usepackage{epsfig}

\begin{document}

\title{Towards characterization of exoplanetary atmospheres 
with~the~VLT~Interferometer}

\author{V. Joergens \and A. Quirrenbach} 
  \institute{Sterrewacht Leiden / Leiden Observatory, 
PO Box 9513, 2300 RA Leiden, The Netherlands }

\maketitle 

\begin{abstract}

The direct observation of extrasolar planets and their spectra
is coming into reach
with the new generation of ground-based near-IR interferometers, like the 
Very Large Telescope Interferometer (VLTI).
The high contrast between star and planet
requires an excellent calibration of atmospheric distortions.
Proposed techniques are the observation of
color-differential or closure phases.
The differential phase, however, is only in a first order 
approximation independent of atmospheric influences
because of dispersion effects. 
This might prevent differential phase observations of extrasolar planets. 
The closure phase, on the other hand, is immune to atmospheric phase errors
and is therefore a promising alternative.
We have modeled the response of the closure phase instrument AMBER at the 
VLTI to realistic models of known extrasolar planetary systems taking 
into account their theoretical spectra as well as the geometry of the VLTI. 
We present a strategy to determine the geometry of the 
planetary system and the spectrum of the extrasolar planet from closure
phase observations in a deterministic way without any a priori assumptions.
We show that the nulls in the closure phase do only depend on
the system geometry but not on the planetary or stellar spectra.
Therefore, the geometry of the system can be determined by measuring
the nulls in the closure phase and braking the remaining ambiguity due to 
the unknown
system orientation by means of observations at different hour angles. 
Based on the known geometry, the planet spectrum can be directly synthesized 
from the closure phases.

\keywords{Extrasolar planets, planetary atmospheres, interferometry,
closure phase, differential phase}
\end{abstract}

\section{Introduction} 
\label{sec:intro}

Since the first extrasolar planet candidate was discovered in 1995
orbiting the solar-like star 51\,Peg (Mayor \& Queloz 1995),
more than 130 extrasolar planets have been detected by radial velocity
surveys (e.g. Mayor et al. 2003 for recent discoveries).
So far, the only observational information on the spectrum of 
an extrasolar planet has been obtained for the transiting system 
HD209458 (e.g. Charbonnneau et al. 2002,
Vidal-Madjar et al. 2004).

   \begin{figure}[h]
   \begin{center}
   \includegraphics[width=9cm]{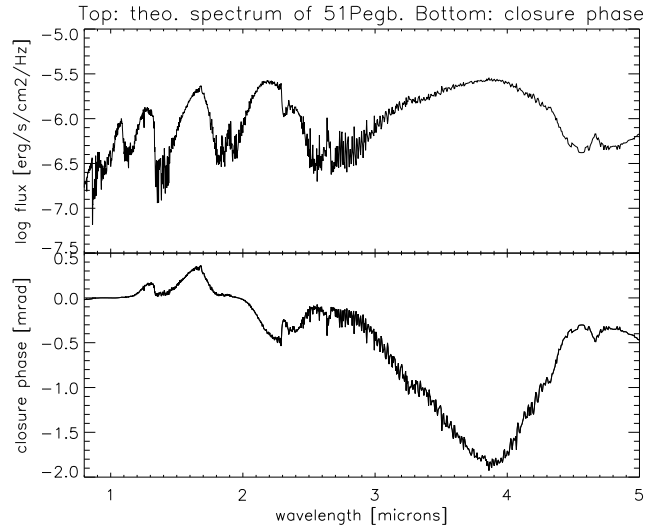}
   \end{center}
   \caption[example] 
   { \label{fig:clphase} 
{\bf Spectral signal of planet in the closure phase.}
\emph{Top panel:} theoretical near-IR spectrum of the
extrasolar planet orbiting 51\,Peg
displaying CO and H$_2$O absorption bands (Sudarsky, Burrows \& Hubeny 2003).
\emph{Bottom panel:} closure phases in milliradian calculated by us
based on the theoretical spectrum of 51\,Peg\,b shown above
as well as on the stellar spectrum for
a simulated observation with the near-IR instrument AMBER/VLTI
using the three telescopes UT1, UT3 and UT4 
(baseline length 102m, 62m, 130m).
It is evident that the spectrally resolved
closure phases contain a wealth of spectral information
of the planet.}
   \end{figure}

In order to gain deeper insights into the physics of extrasolar planets and 
directly detect e.g. their atmospheres, observations with
milliarcsec spatial resolution and at an unprecedented dynamical range
are required (e.g. 10$^{-5}$ in the near-IR for 51\,Peg).
Such observations are coming into reach
with the new generation of ground-based near-IR interferometers, like the 
VLTI.
Observations of high contrast systems, like extrasolar planets,
require an excellent calibration of atmospheric distortions.
It has been proposed to observe extrasolar planets and their spectra 
through the differential phase or closure phase method
(e.g. 
Quirrenbach \& Mariotti 1997, 
Akeson \& Swain 1999, 
Quirrenbach 2000,
Lopez \& Petrov 2000, 
Segransan 2001,
Vannier et al. 2004).
An overview of the mathematical basis for 
the observations of interferometric phases of extrasolar planets
has also been provided by
Meisner (2001) and in particular on the web page referenced 
therein.

Differential phases are measured between different wavelength bands,
which, in a first order approximation, are subject to the same optical 
path length
fluctuations introduced by atmospheric turbulence.
The resulting differential phase errors can therefore be
eliminated in the data reduction to a certain degree.
However, higher order effects due to random dispersion in the atmosphere 
as well as
in the delay lines (caused mainly by variations of the humidity of the air)
can be a significant source of systematic errors and might prevent 
one from reaching
the necessary precision to observe differential phases of extrasolar planets. 

A more promising approach is the observation of closure phases, 
which are measured for a closed chain of three or more telescopes.
The closure phase $\Phi_\mathrm{cl}$
is the sum of the individual single-baseline phases
$\Phi_{ij}$ measured on the baseline between telescope $i$ and $j$
of the array:
\begin{eqnarray*}
\Phi_\mathrm{cl} = \Phi_{12} + \Phi_{23} + \Phi_{31}  .
\end{eqnarray*}

The closure phase is independent of atmospheric phase errors and 
allows therefore a self-calibration, which was first recognized by 
Jennison (1958). 

Closure phase observations from the ground will be possible with 
the AMBER (Astronomical Multi-BEam combineR) instrument at the 
VLTI operated by ESO (Petrov et al. 2003, Malbet et al. 2004) .
AMBER is 
operating in the near-IR J, H, and K bands from 1.1 to 2.4$\mu$m.
It is currently undergoing commissioning at Cerro Paranal and 
is scheduled to be available for regular observations in 2005.
With a multi-beam combiner element and 
a fringe dispersing mode, it has the
capability of combining the light of three
telescopes (UTs or ATs), and performing spectrally resolved 
differential and closure phase observations.
There are three different spectral resolutions available:
low (R=$\lambda$/$\Delta \lambda$=35), medium 
(R=500\dots1000) and high (R=10000\dots15000).

We have modeled the response of the AMBER instrument at the VLTI for known
extrasolar planetary systems taking into account their theoretical
spectra as well as the geometry of the VLTI. 
The closure phase depends not only on the contrast ratio between 
planet and star but also on the (generally unknown) system geometry
as well as on the interferometer geometry and the hour angle of observation.
In the following, we present a method
to determine the planetary system geometry and the 
planet spectrum from closure phase observations in a deterministic way
without any a priori assumptions.

   \begin{figure}[t]
   \begin{center}
   \includegraphics[width=7.2cm,angle=90]{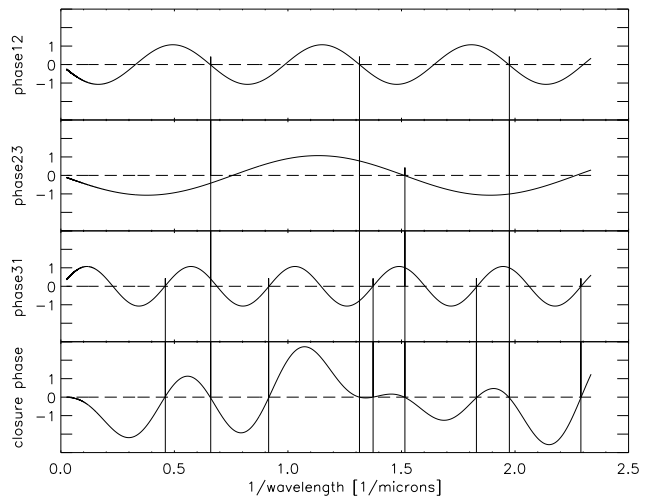}
   \end{center}
   \caption[example]
   { \label{fig:nulls}
   {\bf Relation between the closure phase and the 
   corresponding single-baseline phases.}
   The figure shows simulated phase measurements
   for a system comprised of two point sources with a 3 milliarcsec separation 
   and a constant flux ratio of the components
   of 10$^{-3}$ observed at zenith
   with the VLTI telescopes UT1, UT3 and UT4.
   The three top panels display the phases calculated
   for the single-baselines B$_{12}$, B$_{23}$ and 
   B$_{31}$. (We note that they cannot be measured without
   a phase reference (e.g. PRIMA) and are displayed here only for 
   theoretical considerations.)
   The bottom panel shows the simulated closure phase for
   these three telescopes. All phases are in millirad.  
   It is evident that the nulls in the 
   closure phase are always also nulls in 
   the single-baselines phases. In particular, every second
   null of a single-baseline phase is also a null in the 
   closure phase. 
   }
   \end{figure}

\section{Modeling of closure phases of exoplanets
}

We have written a code to simulate closure phase observations
with AMBER
for known extrasolar planetary systems, taking into account their
theoretical spectra as well as the geometry of the VLTI.
We use theoretical spectra, which have been calculated by 
Sudarsky, Burrows \& Hubeny 2003 (see their paper 
for details). 
For the separation between star and planet the published semi-major axis
is taken, whereas the orientation of the system is unknown and an
arbitrary value was chosen for the purpose of the simulations.

Fig.\ref{fig:clphase} 
displays in the top panel a theoretical spectrum of the giant irradiated 
planet orbiting the solar-like star 51\,Peg.
The bottom panel of Fig.\,\ref{fig:clphase} shows the modeled 
closure phase response 
to the planetary system 51\,Peg in a simulated observation 
with AMBER and the three 8\,m 
unit telescopes UT1, UT3 and UT4 (baseline lengths 102m, 62m, 130m).

It can be seen from Fig.\ref{fig:clphase} that the spectrally resolved
closure phases contain information on the planetary spectrum.
However, they also depend
on the interferometer geometry, the hour angle of the observations,
the spectra of the star, and on the planetary system
geometry.
The latter is generally unknown for radial velocity 
planets. 

Further results of the simulations are shown in Fig.\ref{fig:nulls} and
Fig.\ref{fig:rotsynt} and are discussed in the following sections.

\section{Nulls in the closure phase}

We show in this section that the nulls in the closure phase are independent
of the planet\,/\,star contrast ratio but do only depend on the
planetary system geometry,
i.e. separation and orientation of the system. 
Therefore, the measurement of the 
wavelength for which the closure phase is zero, is
the first step in disentangling the 
planet spectrum from the closure phase.

In a mathematical sense, the condition for single-base\-line phases $\Phi_{ij}$
to be zero is that the dot product
between separation vector $\vec{s}$ (pointing from the star to the planet) 
and the corresponding projected baseline
$\vec{B_{ij}}$ equals multiples of $\lambda/2$:

\begin{equation}
\label{equ:phasenull}
\vec{s}*\vec{B_{ij}}     = n \cdot \lambda/2  \quad \Longrightarrow \quad
\Phi_{ij} = 0  .
\end{equation}

Thus, the nulls in the single-baseline phase depend only on the
interferometer geometry $\vec{B_{ij}}$ and the planetary 
system geometry $\vec{s}$, 
but are independent of the planet and star spectra.
(We note that
the absolute interferometric phase measured on a single-baseline is not
an accessible quantity without phase referencing, as planned for example 
for the instrument PRIMA, which is under development for the VLTI.)

We have shown analytically (Joergens \& Quirrenbach 2004)
that there is also a simple condition for nulls in the closure phase:
the closure phase is zero if
the dot product between 
separation vector $\vec{s}$ and one of the involved baseline vectors 
$\vec{B_{ij}}$ equals multiples of $\lambda$:

\begin{eqnarray}
\label{equ:clphasenull}
\vec{s}*\vec{B_{ij}}     = n \cdot \lambda  \quad \Longleftrightarrow \quad 
 |s| \cdot |B_{ij}| ~ \vartheta_{ij} = n \cdot \lambda \\
\Longrightarrow \quad \Phi_\mathrm{cl} = \sum_{ij} \Phi_{ij} = 0
\end{eqnarray}

with $\vartheta_{ij}$ being the angle between $\vec{s}$ and $\vec{B_{ij}}$.

Comparison with Eqn.\,\ref{equ:phasenull}
shows that for every second null in a single-baseline phase, 
the closure phase has also a null.

This is illustrated by 
Fig.\ref{fig:nulls}, which displays the simulated single-baseline
as well as closure phases
for two point sources with a constant contrast ratio
of 10$^{-3}$ for all wavelengths. Considering such a 'flat spectrum'
allows us to show the relation between
the closure phase and the corresponding individual phases
without the complication of the complex spectral features
of the source.
It can be seen from the plot that the nulls in the closure phase
are always also nulls in one of the corresponding single-baseline phases.
Furthermore, every second null of a single-baseline phase is
also a null in the closure phase.

This shows that the nulls in the closure phase are also independent of 
the planet\,/\,star contrast ratio and do only depend on the 
system geometry.

\section{Earth rotational synthesis}

We now proceed to describe an algorithm to determine the spectrum of a 
planet and the geometry of the 
star-planet system (i.e. the angular separation and position angle of the 
planet with respect to the star at the time of observation) from closure 
phase observations, without any a priori knowledge of these quantities. 
The stellar spectrum can be presumed to be known, however, since this
is easily measurable with a simple spectrograph.
Determining the planet spectrum is therefore equivalent to measuring the 
planet\,/\,star contrast ratio.

   \begin{figure}[t]
   \begin{center}
   \includegraphics[height=9cm]{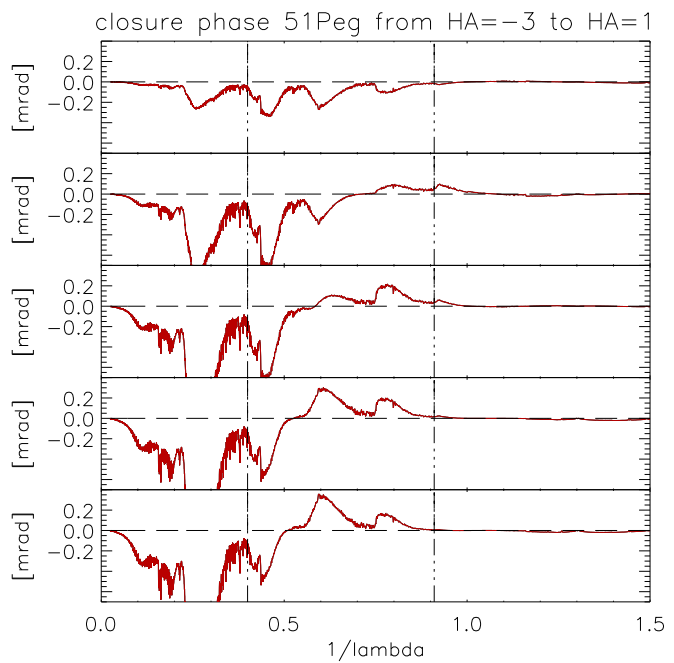}
   \end{center}
   \caption[example] 
   { \label{fig:rotsynt}
    {\bf Closure phase time series for the planet system 51\,Peg.}
   Simulations for observations with the VLTI array UT1, UT3, UT4
   for hour angles HA\,=\,-3, -2, -1, 0, +1\,hr (from top to bottom).
   Vertical lines indicate the wavelength range covered by AMBER
   (the near-IR bands between 1.1\,$\mu$m and 2.5\,$\mu$m).
   }
   \end{figure} 

   \begin{figure*}[t]
   \begin{center}
   \begin{tabular}{cc}
   \includegraphics[width=7.5cm]{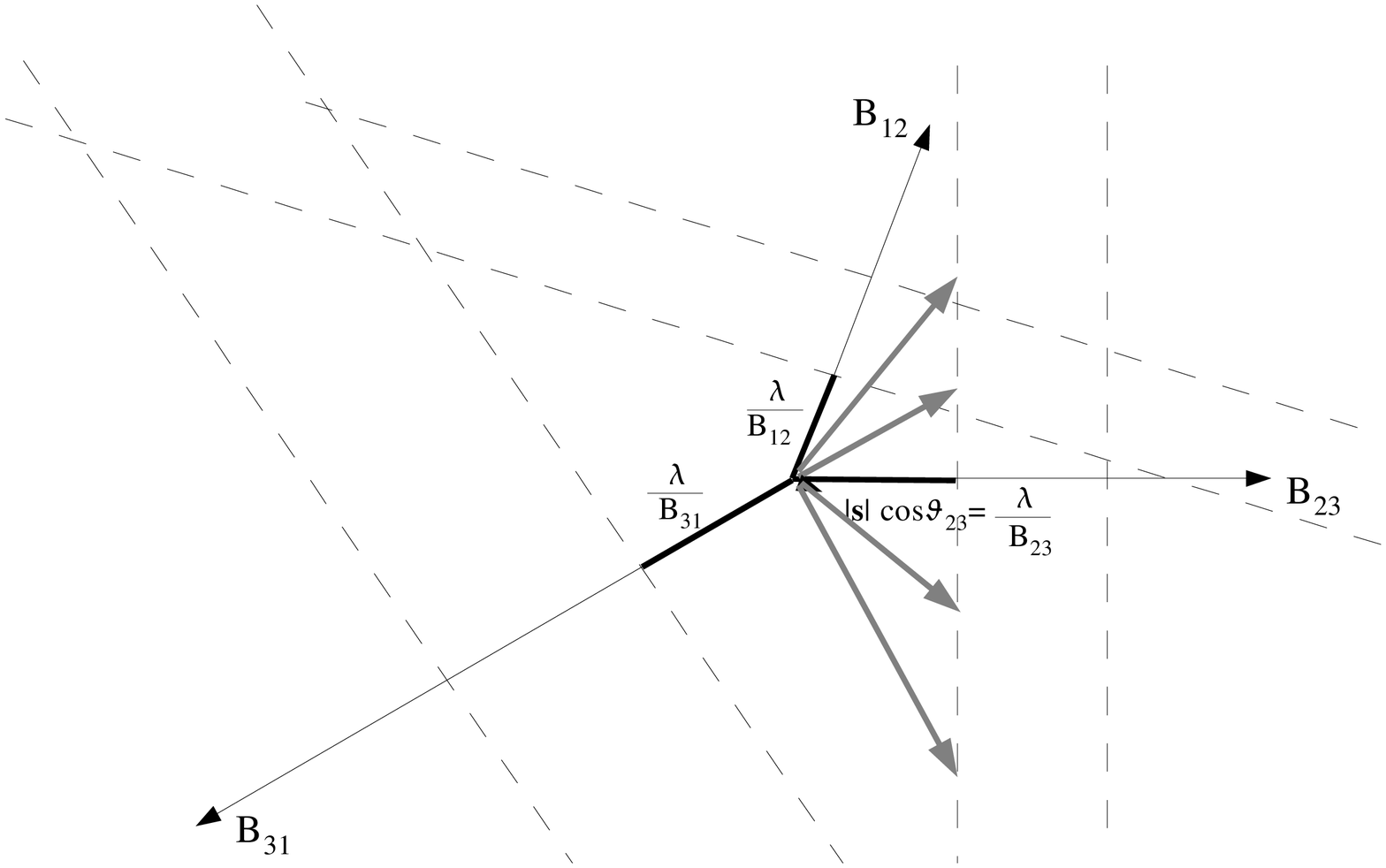}
   \hspace{1cm}
   \includegraphics[width=7.5cm]{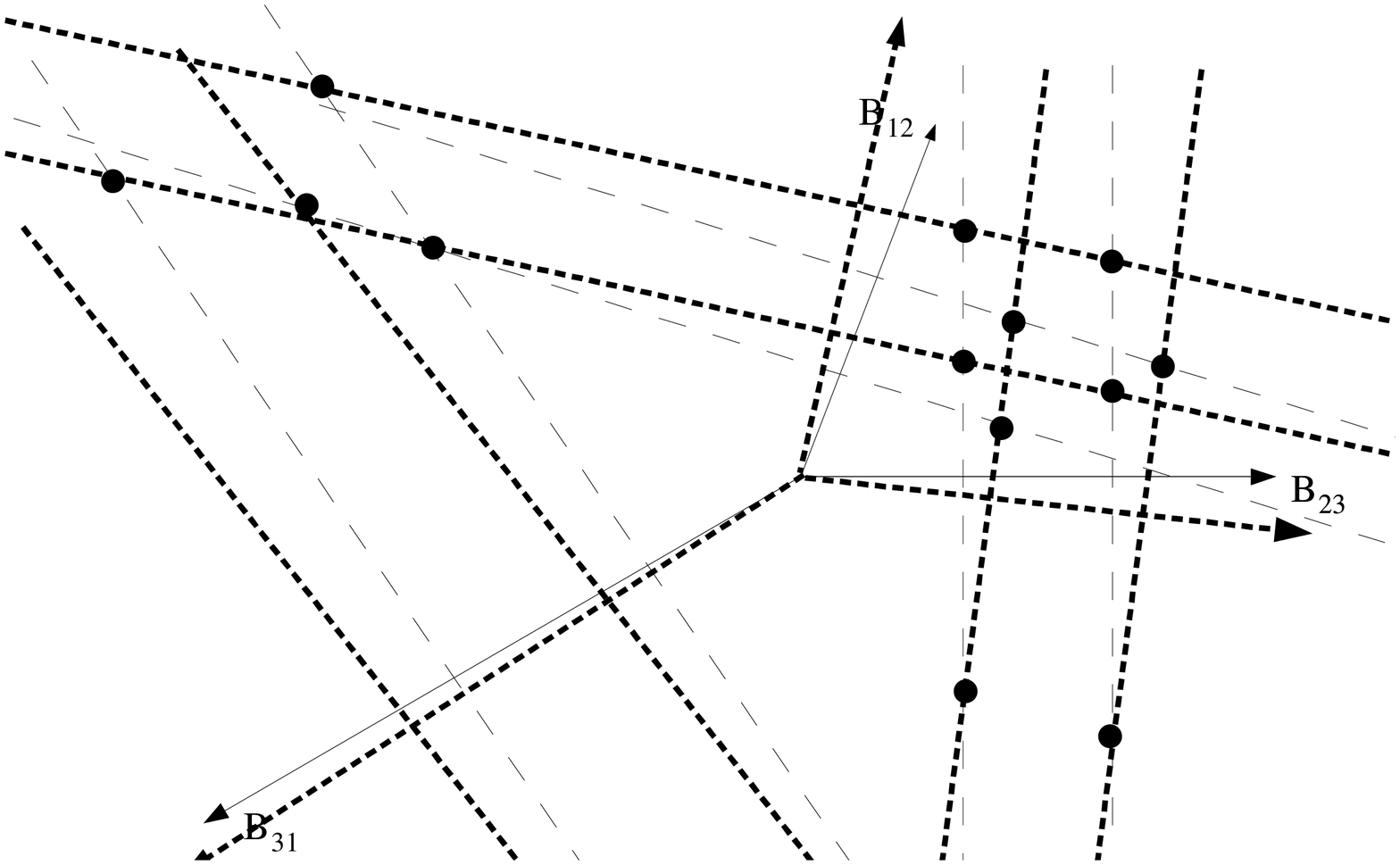}
   \end{tabular}
   \end{center}
   \caption[example] 
   { \label{fig:liniennetz}
   {\bf Possible geometries of a star\,--\,planet system as derived 
   from nulls of the closure phase.}
   \emph{Left:} Determination of the nulls in the closure phase
   signal of a planetary system at a certain hour angle is a measure for
   the projection of the separation vector $\vec{s}$ between
   star and planet onto one of the three projected baselines.
   Due to the unknown orientation $\vartheta_{ij}$ we do not know onto
   which one. The dashed lines indicate the possible locations of $\vec{s}$
   for the nulls corresponding to $n=1$ and 2.
   \emph{Right:} Observations at a later hour angle gives another set of lines.
   The possible locations for $\vec{s}$ are now reduced to several grid points
   as marked by the intersections with the former lines. With a third
   observation, it will be possible to select the correct intersection point.
   }
   \end{figure*}

As described in the previous section, the nulls of the closure phase 
contain information about the system geometry because they are related to 
the nulls of the single-baseline phases. However, it is not known a priori 
to which single baselines the individual nulls in the 
closure phase correspond. 
We can thus interpret Eqn.\,\ref{equ:clphasenull} as follows:
By determining a wavelength for which the closure phase is zero, we have 
a measure for the projection $|\vec{s}| \cdot \vartheta_{ij}$ 
of the separation vector $\vec{s}$ onto one of the individual baselines
$\vec{B_{ij}}$ but we do not know onto which one.
Furthermore, we do not necessarily know the order $n$ of the null.
The geometrical locus of all possibilities
for $\vec{s}$ is therefore a set of straight lines, as shown in the 
left panel of Fig.\,\ref{fig:liniennetz}. Each line in this figure is 
perpendicular to one of the baselines and corresponds to an assumed set of 
{$i,j,n$}. For clarity only the lines corresponding to $n$ = 1 and $n$ = 2 
are shown.

The set of projected baseline vectors formed by the three telescopes 
changes with time due to Earth's rotation. Therefore, a second observation
gives an independent constraint on the geometry as shown in the right panel
of Fig.\,\ref{fig:liniennetz}. The possible system geometries are now the intersections of the set of lines corresponding to the first observation with the set
corresponding to the second. This discrete set of points has been marked
with filled circles in the figure. A third observation will produce yet another
independent set of lines. In general, this set will pass through only one of 
the previously marked points: this is the true separation vector.
It is thus possible to derive the system geometry unambiguously with three 
observations.

Once the separation vector is known, it is straightforward to determine
the planet\,/\,star contrast ratio for each wavelength 
through a numerical inversion of the expression for the closure phase.
The only difficulty occurs at or very close to the nulls, where the 
signal-to-noise is zero or very low.
However, since three observations are needed in any case for the determination
of $\vec{s}$, one can perform the three 
inversions and
combine them with (wavelength-dependent) weights appropriate for their
respective signal-to-noise ratios. Consider, for example, the wavelength range
around $\lambda = 1.67\,\mu$m ($1 / \lambda = 0.6\,\mu$m$^{-1}$) in
Fig.~\ref{fig:rotsynt}. The observation at $-1$\,hr (third panel) gives very
low signal-to-noise close to the null near $1 / \lambda = 0.6\,\mu$m$^{-1}$,
but observations at other times can be used to infer the planet spectrum in
this wavelength region.

\section{Outlook}

We have shown that the separation vector and spectrum of extrasolar planets can
be determined from closure phase measurements in a deterministic way, with a
non-iterative algorithm, and without any a priori assumptions. For a practical
application of this technique, a few additional complications will have to be
considered:
\begin{itemize}
\item{The assumption that the star and the planet are point sources will have
to be relaxed. Taking into account that the star is slightly resolved by the
interferometer complicates the analysis and implies that
the relation between nulls in the closure phase and the single-baseline phases
is only approximately fulfilled.}
\item{For observations from the ground, the useful wavelength range in the
near-infrared is limited to the atmospheric windows and thus non-contiguous.}
\item{The finite signal-to-noise ratio of realistic observations will lead to
an uncertainty in determining the exact wavelengths of closure phase nulls. The
lines and intersection points of Fig.~\ref{fig:liniennetz} will thus be
broadened.}
\item{The non-zero time needed to accumulate sufficient signal-to-noise on the
closure phase means that the interferometer geometry will change slightly
during the observations. This is again equivalent to a slight broadening of the
allowed regions in Fig.~\ref{fig:liniennetz}.}
\item{If the time required to accumulate all observations is not short compared
to the orbital period of the planet, the motion of the planet will also
contribute an uncertainty in the derived geometry.}
\end{itemize}
These practical complications will certainly be at least partially compensated
by a larger number of observations; one would probably take ten or more rather
than the minimum three. One could also combine the algorithm presented here
with a global $\chi^2$ minimization, in which the orbital parameters and
spectrum of the planet are fitted to the observations. The purpose of our
algorithm would then consist of providing a robust starting point for the
$\chi^2$ minimization algorithm, which is usually crucial to ensure convergence
to the correct minimum.

The AMBER instrument has arrived on Cerro Paranal this year and the first
commissioning runs have taken place. It is now necessary to determine if the
required closure phase precision (better than 0.1\,mrad, see
Fig.~\ref{fig:clphase}) can be reached. If so, the VLTI will provide
unprecedented opportunities for observations of extrasolar planets and their
spectra.

\begin{acknowledgements}

We are grateful to our colleagues at the Sterrewacht Leiden, Jeff Meisner,
Bob Tubbs and Walter Jaffe for 
helpful discussions on the topic of this article.
VJ acknowledges support by a Marie Curie Fellowship of the
European Community programme 'Structuring the European Research Area'
under contract number FP6-501875.

\end{acknowledgements}


\end{document}